\documentclass[11pt,a4paper]{article}
\usepackage{jheppub}
\usepackage[T1]{fontenc}

\allowdisplaybreaks
\title{A renormalizable supersymmetric SO(10) model with natural doublet-triplet splitting}
\author{Ying-Kang Chen}
\author{and Da-Xin Zhang}
\affiliation{School of Physics and State Key Laboratory of Nuclear Physics and Technology, \\
Peking University, Beijing 100871, China}

\emailAdd{ykchen@pku.edu.cn}
\emailAdd{dxzhang@pku.edu.cn}

\abstract{
We propose a renormalizable supersymmetric SO(10) model where the doublet-triplet splitting
problem is solved using the Dimopoulos-Wilczek mechanism.
An unwanted coupling is forbidden  through a filter sector.
To suppress proton decay without spoiling
gauge coupling unification, there is a problem in the weak doublets which requires further improvements.}
\begin{document}
\maketitle
\flushbottom
\pagenumbering{arabic}

\newcommand{\0}{SO(10)}
\newcommand{\5}{SU(5)}
\newcommand{\3}{$SU(3)_C$}
\newcommand{\2}{$SU(2)_L$}
\newcommand{\1}{$U(1)_Y$}
\newcommand{\Group}{$\textrm{SU(3)}_C\times\textrm{SU(2)}_L\times \textrm{U(1)}_Y$}
\newcommand{\G}{$G_{321}$}
\newcommand{\group}{$SU(3)_C\times U(1)_{\textrm{em}}$}

\newcommand{\ii}{\textrm{i}}
\newcommand{\dd}{\textrm{d}}
\newcommand{\vev}[1]{\langle#1\rangle}
\newcommand{\ran}[1]{|#1\rangle}

\newcommand{\pair}{~\textbf{126}+\overline{\textbf{126}}~}
\newcommand{\del}{$\bm{126}$}
\newcommand{\delb}{$\overline{\bm{126}}$}
\newcommand{\sg}{\Sigma}
\newcommand{\sgb}{\overline{\Sigma}}
\newcommand{\dl}{\Delta}
\newcommand{\dlb}{\overline{\Delta}}

\newcommand{\be}{\begin{equation}}
\newcommand{\ee}{\end{equation}}
\newcommand{\bea}{\begin{eqnarray}}
\newcommand{\eea}{\end{eqnarray}}

\newcommand{\bref}[1]{(\ref{#1})}
\newcommand{\phd}[1]{\Phi^{(D)}_{#1}}
\newcommand{\pht}[1]{\Phi_{#1}}

\section{Introduction}

Grand Unified Theories (GUTs) are very important
candidates for the physics beyond the Standard Model (SM).
In GUT models Higgs fields are unified into complete multiplets of the GUT symmetry.
Consequently, a serious problem in GUT models is that while the weak doublets are massless
compared to the GUT scale,
the color-triplets in the same  multiplets are very heavy as required by proton decay experiments.
This problem is eased in the supersymmetric (SUSY) versions of GUT models where
if this  doublet-triplet splitting (DTS) is realized at tree level,
the non-renormalization theorem protects the DTS from radiative corrections.

In SUSY SO(10) models\cite{clark,moha},
to solve the DTS problem several mechanisms have been developed\cite{mpm,mpm2,mpm3,dw,dw2,dwAAA,dwAAp,lee,dw3,dw32,dw33,dw34,dw35,dw36,dw37,dw38},
among which the Missing Vacuum Expectation Value (Missing VEV) or the Dimopoulos-Wilczek (DW) mechanism\cite{dw,dw2,dwAAA,dwAAp,lee,dw3,dw32,dw33,dw34,dw35,dw36,dw37,dw38}
is widely used.
It uses the adjoint Higgs superfield in  $\mathbf{45}$, usually assisted by Higgs in $\mathbf{54}$,
to break $SO(10)$ down to $SU(3)_c  \times SU(2)_L \times U(1)_R \times U(1)_{B-L}$ (or to
$SU(3)_c  \times SU(2)_L \times SU(2)_R \times U(1)_{B-L}$ when the DW mechanism applies).
Since the $\mathbf{45}$ is anti-symmetric,
two different Higgs in $\mathbf{10}$ ($\mathbf{10_{1,2}}$)are needed to couple with $\mathbf{45}$.
When the $\mathbf{45}$ acquires a VEV only in the $(15,1,1)$ but not in the $(1,1,3)$ direction,
where the numbers in the brackets are representations under the $SU(4)_C\times SU(2)_L\times SU(2)_R$
subgroup of SO(10),
all the color-triplets are massive while the weak doublets can be massless.

To further break down the symmetry into the SM gauge group,
representations containing $B-L$ breaking VEVs are needed.
In the non-renormalizable models, Higgs in $\mathbf{16+\overline{16}}$ are introduced.
This will bring into the dangerous dimension-4 $R$-parity violating operators
which lead to too rapid proton decay rates.
Also, to describe correct fermion masses, operators of higher dimensions are used which make
the models have less predictive abilities.
In the renormalizable models,
we need Higgs in $\mathbf{126+\overline{126}}$\cite{ferm1},
usually even accompanied by
$\mathbf{120}$\cite{Berto,ferm20,Aulakh,ferm2,ferm21,Lavoura,Grimus,Aulakh2,Joshipura,ferm3,Joshipura2,ferm4},
to describe fermion masses.
The DW mechanism is mostly realized in the non-renormalizable models\cite{dw3,dw32,dw33,dw34,dw35,dw36,dw37,dw38},
except in \cite{lee} which is  (partially) renormalizable.

In the renormalizable model of \cite{lee},
the key point is to introduce the couplings $\mathbf{\overline{126} \cdot 210 \cdot 126}$ and
$\mathbf{10_1\cdot 210\cdot 126}$ while the coupling
$\mathbf{10_1\cdot 210\cdot \overline{126}}$ is absent.
This is realized by enforcing an extra symmetry.
An unsatisfactory aspect in the model of \cite{lee} is the introduction of explicit symmetry
breaking terms in the model.
Also, the transformation properties of $\mathbf{10_1}$ and $\mathbf{\overline{126}}$
are different under this extra symmetry,
so that they cannot couple with the matter fields at the same time.
Then the non-renormalizble Yukawa couplings
are inevitable  to account for correct fermion masses.
Furthermore, $\mathbf{120}$ cannot be incorporated easily in the model since the coupling
$\mathbf{120\cdot 45\cdot 10_2}$, which breaks the DW mechanism, generally exists,
if the $\mathbf{120}$ has the same transformation property as the $\mathbf{10_1}$
under the extra symmetry.

In this paper,
our aim is limited to present a renormalizable model to realize the DW mechanism.
Comparing with  \cite{lee} in which the only renormalizable Yukawa couplings are those with Higgs in $\mathbf{10_1}$,
we can include renormalizable Yukawa couplings with  $\mathbf{\overline{126}}$  and even with $\mathbf{120}$,
so that fermion masses are fully realistic.
However, we will not fully solve the problem on proton decay
which needs mechanism\cite{dlz2} other than the DW mechanism,
although the proton decay amplitudes mediated by the color triplets  in $\mathbf{10_1}$ are suppressed
without spoiling gauge coupling unification.
We will first present a renormalizable model without $\mathbf{120}$,
next include $\mathbf{120}$ resulting a pair of massless doublets and a pair of massless triplets,
then propose a method to give masses to the triplets using the DW mechanism by introducing a ``filter sector''
to forbid the unwanted $\mathbf{120\cdot 45\cdot 10_2}$ coupling,
and give a full model. We will finally summarize.

\section{A model without $\mathbf{120}$}

In order to clarify how to implement the DW mechanism in renormalizable models to realize the DTS,
we present a model without $\mathbf{120}$ firstly.
The field content consists three $\mathbf{45}$ ($A, A^\prime, A^{\prime\prime}$),
two $\mathbf{54}$ ($S$, $S^\prime$),
one pair of $\mathbf{126}+\overline{\mathbf{126}}$ ($\Delta$ and $\overline{\Delta}$),
one $\mathbf{210}$ ($\Phi$), and
two $\mathbf{10}$s ($H_{1,2}$). 
One SO(10) singlet ($Y$) is also introduced whose VEV acts as
the mass for the first two $\mathbf{45}$s ($A, A^\prime$) which are to acquire DW VEVs.
The third $\mathbf{45}$ ($A^{\prime\prime}$) is used to eliminate redundant Goldstones [8,9].
The first $\mathbf{54}$ ($S$) is used to generate the DW VEVs,
while the second $\mathbf{54}$ ($S^\prime$) is required to maintain SUSY.
$\Delta$ and $\overline{\Delta}$ are used to reduce the rank of gauge group.
$\mathbf{210}$ is used in mixing $H_1$ and $\Delta$.

We impose a $\mathbb{Z}_{12}$ symmetry. Transformation properties of all
the Higgs superfields  are listed in Table I,
besides the matter superfields of all the three generations whose $\mathbb{Z}_{12}$ charges are taken as $-1$.
The most general Higgs superpotential is
\begin{eqnarray}
W&=&W_{DW}+W_{SB}+W_{DT},\nonumber
\end{eqnarray}
where
\begin{eqnarray}
W_{DW}&=&Y A A^\prime+SA A^\prime+AA^\prime A^{\prime\prime},\label{WDW}\\
W_{SB}&=& \frac{1}{2}m_\Phi \Phi^2 + \Phi^3
+\Phi A^{\prime\prime 2}+\frac{1}{2}m_{A^{\prime\prime}}A^{\prime\prime 2}
+\frac{1}{2}m_YY^2\nonumber\\
&+&\frac{1}{2}m_{S} S^{2}
+\frac{1}{2}m_{S^\prime} S^{\prime 2}+S^{\prime 3}+S^2 S^{\prime}+S^{\prime} \Phi^2\nonumber\\
&+&S^{\prime} A^{\prime\prime 2}
+S S^{\prime} A^{\prime\prime}+S S^{\prime} Y,\label{WSB}\\
W_{DT}&=&\frac{1}{2} m_2 H_2^2 + S^{\prime} H_2^2 + H_1 A H_2 + (H_1 + \bar\Delta) \Phi \Delta+m_\Delta \Delta \overline{\Delta}.
\label{WDT}
\end{eqnarray}
Here we have suppressed all dimensionless couplings for simplicity.
In (\ref{WDT}) we do not include
the nonrenormalizable couplings such as $\frac{1}{\Lambda}H_1H_1AA$ which will otherwise
break the DW mechanism of realizing the DTS.
In the nonrenormalizable models, such higher dimensional couplings must be considered
because other couplings of the same properties are used.
However, in the renormalizable models, we can simply
regard whether these couplings exist or not depend on if there are states of suitable representations
which can generate the relevant tree-level diagrams.
Here, loop diagrams are not important, as they are SUSY breaking effects which are small.
Note that even in the nonrenormalizable models, how large the higher dimensional couplings
are is not known \cite{dwAAp}.

In $W_{DW}$, the first two terms are the modified form for realization of the DW mechanism\cite{dwAAp},
while the third term, which has no effect on the F-flatness conditions of keeping SUSY at high energy,
is introduced to eliminate extra Goldstone particles in $A$ and $A^\prime$\cite{dwAAA}.

\begin{table}\begin{center}
\begin{tabular}{|c|c|c|c|c|c|c|c|c|c|c|c|}
  \hline
    & $H_1$ & $H_2$ & $\overline{\Delta}$ & $\Delta$ & $\Phi$ & $A$  & $A^\prime$ & $A^{\prime\prime}$& $S$&$S^\prime$ &$Y$\\
  \hline
  $\mathbb{Z}_{12}$ & 2 & 6 & 2 & -2 & 0 & 4 & 2 & 6 &  6&0 & 6\\
\hline
\end{tabular}\label{tab1}\end{center}
\caption{A renormalizable model without $\mathbf{120}$ with a $Z_{12}$ symmetry imposed.
A field $X$ with charge $q$ transforms as $X\rightarrow X e^{i q\pi/{6}}$.}
\end{table}

Under $SU(4)_C\otimes SU(2)_L\otimes SU(2)_R$, there are possibly two  VEVs  for every $\mathbf{45}$,
 {\it e.g.} ${A_1}$ and ${A_2}$ for $A$  in the  $(1,1,3)$ and $(15,1,1)$ directions, respectively.
Setting the F-terms of $A_1^\prime$ and $A_2^\prime$ to zeros,
we get the following equations
\begin{eqnarray}
0=A_1\left(Y\text{  }+\frac{3}{2\sqrt{15}} S \right), ~
0=A_2\left(Y\text{  }-\frac{1 }{\sqrt{15}}\text{  }S\right).\nonumber
\end{eqnarray}
Hereafter we will use the same symbols for the fields and their VEVs.
One set of the solutions are
\begin{eqnarray}
A_1=0, ~~A_2\neq 0, ~~Y=\frac{1 }{\sqrt{15}}\text{  }S,
\end{eqnarray}
which realize the DW VEVs.
It also follows $A_1^\prime=0$ and $A_2^\prime\neq 0$ when the F-flatness conditions for $A$ are enforced.

Notice that using the DW VEV, $A_1=0$,
the doublets in $H_2$ separate from the other Higgs doublets .
Then the up-type doublet in $\Delta$ acquires masses only
through its couplings with the down-type doublets in
$H_1$ and in $\overline{\Delta}$.
To be explicitly,
the mass matrix for the weak doublets is
\begin{eqnarray}\label{mass0}
\left(
\begin{array}{c|c}
0_{2\times 2}&\begin{array}{cc}A_{11} &0 \\ A_{21}&0\end{array}\\
\hline
B_{2\times 2}&C_{2\times 2}
\end{array}
\right),
\end{eqnarray}
where the columns are ordered as
($H_1^u,\overline{\Delta}^u; {\Delta}^u,{\Phi}^u$)
and the rows are ($H_1^d,\overline{\Delta}^d; {\Delta}^d,{\Phi}^d$).
Expressions for individual elements can be found in \cite{msso10,msso102}.

(\ref{mass0}) has  a zero eigenvalue.
To be more explicitly, a combination of the first two rows gives a new row with all its elements to be zeros,
and the corresponding  combination of $H_1^d$ and $\overline{\Delta}^d$ is identified as the $H_d^0$ of the
Minimal Supersymmetric Standard Model (MSSM).
There also exists  a massless up-type Higgs doublet $H_u^0$ of the MSSM,
since the first, the second and the last columns in (\ref{mass0}) can give a column with all its elements to be zeros
so that the massless $H_u^0$  has components from $H_1^u,\overline{\Delta}^u$ and ${\Phi}^u$.
The above arguments do not apply for the color-triplet sector due to the coupling
$H_1 A H_2$ where $A_2\neq 0$ couples  these triplets so that $H_2$ does not separate from other Higgs
in the triplet sector.

The above observation follows \cite{lee} directly.
The key point is the presence of  $H_1\Delta \Phi$ and the absence of $H_1\overline\Delta \Phi$,
otherwise two or null zero eigenvalues exist in the doublet mass matrix.
The differences from \cite{lee} are that
the present model is fully consistent with the imposed symmetry,
and $H_1$ and $\overline{\Delta}$ have the same transformation properties under this symmetry
so that they can couple with matter fields simultaneously.

\section{The massless states in a model with  $\mathbf{120}$}

In the presence of $\mathbf{10, 120, \overline{126}, 126, 210}$,
we need to construct a model which contains a pair of massless weak doublets and a pair
of massless color triplets while keeping all other states massive.
After realizing the DW mechanism which will be performed in the next section,
the color triplets will get masses while the weak doublets keep massless.

We will introduce two $\mathbf{120}$ ($D_{1,2}$) and the relevant  superpotential is
\begin{eqnarray}
(H_1+D_1+\overline{\Delta})\Phi (\Delta+D_2 )
+ m_D D_1 D_2 +m_\Delta \Delta\bar\Delta + \frac{1}{2} m_\Phi \Phi^2 + \Phi^3.\nonumber\\\label{masslessW}
\end{eqnarray}
Note from (\ref{masslessW}) all the states in $D_{1,2}, \overline{\Delta},\Delta,\Phi$ get masses.
The only possible massless states are the doublets and triplets in $10_1$.
Now we exam if there are still massless states in the presence of the interactions  (\ref{masslessW}) which
mix all the doublets and triplets.

The contents of all doublets, besides the pair from $H_2$ which  decouple,  are
\begin{itemize}
  \item one pair from $H_1$ ($H_1^u+H_1^d$);
  \item one pair from $\overline{\Delta}$ ($\overline{\Delta}^u+\overline{\Delta}^d$);
  \item one pair from ${\Delta}$ (${\Delta}^u+{\Delta}^d$);
  \item two pairs from each of $D_{i=1,2}$ ($D_i^u+D_i^d$ and $D_i^{\prime u}+D_i^{\prime d}$). The unprimed and primed doublets are in (1,2,2) and (15,2,2) under $SU(4)_C\otimes SU(2)_L\otimes SU(2)_R$, respectively;
  \item one pair from $\Phi$ (${\Phi}^u+{\Phi}^d$).
\end{itemize}
From (\ref{masslessW}),
the mass matrix for the doublets is
\begin{eqnarray}\label{massD}
\left(
\begin{array}{c|c}
0_{4\times 4}&\begin{array}{cc}A_{4\times3} &0_{4\times 1}\end{array}\\
\hline
B_{4\times 4}&C_{4\times 4}
\end{array}
\right),
\end{eqnarray}
where the columns are ordered as
($H_1^u,D_1^u,D_1^{\prime u},\overline{\Delta}^u; {\Delta}^u,D_2^u,D_2^{\prime u},{\Phi}^u$)
and the rows by replacing the superscripts ``$^u$'' by ``$^d$''.
The matrices $B$ is fully ranked,  $C$ is not fully null  while most of elements in $A$ are nonzero.
The key ingredient  is the $0_{4\times 1}$ sub-matrix as a consequence of
the absence of the coupling $H_1 \Phi\overline{\Delta}$ which, if present, will induce a nonzero $4\times 1$ matrix
with elements proportional to the VEV of $\overline{\Delta}$.
It is clear that there is a zero eigenvalue in (\ref{massD});
or more explicitly,
since a combination of the first four rows in  (\ref{massD}) can give a row with all its
elements to be zeros,
a combination of $H_1^d,D_1^d,D_1^{\prime d},\overline{\Delta}^d$ gives the massless eigenstate $H_d^0$
of the MSSM.
Similarly, the massless eigenstate $H_u^0$ has components from  the doublets in $H_1^u,D_1^u,D_1^{\prime u},\overline{\Delta}^u$ and ${\Phi}^u$.

For the triplets, there are one more pair compared to the doublets.
They are one more color-triplet from $\overline{\Delta}$ and one more color-anti-triplet from ${\Delta}$.
Ordering the columns as
($H_1^T,D_1^T,D_1^{\prime T},\overline{\Delta}^T,\overline{\Delta}^{\prime T};
{\Delta}^T,D_2^T,D_2^{\prime T},{\Phi}^T$) and the rows as
($H_1^{\overline T}, D_1^{\overline T}, D_1^{\prime {\overline T}}, \overline{\Delta}^{\overline T};
{\Delta}^{\overline T}, \Delta^{\prime {\overline T}}, D_2^{\overline T},
D_2^{\prime {\overline T}}, {\Phi}^{\overline T}$),
the mass matrix for the triplets is
\begin{eqnarray}\label{massT}
\left(
\begin{array}{c|c}
0_{4\times 5}&\begin{array}{cc}A_{4\times3} &0_{4\times 1}\end{array}\\
\hline
B_{5\times 5}&C_{5\times 4}
\end{array}
\right),
\end{eqnarray}
again, there is a pair of massless color triplets which contain
nonzero components from $H_1^T$ and $H_1^{\overline T}$, respectively.
In the next section, this pair of triplets will be given masses through the DW mechanism
while the doublets are keeping massless.

\section{The filter sector for DW mechanism in the presence of  $\mathbf{120}$}

In renormazible SUSY SO(10) models, usually a Higgs in  $\mathbf{120}$  is also
needed to couple with the fermion sector through Yukawa interactions\cite{ferm2,ferm3,ferm4}.
This requires that $D_1$ has the same symmetry property as
$H_1$ and ${\overline{\Delta}}$.
When the triplets of  $H_1$ receive masses through the coupling
$H_1 A H_2$, the DW VEV $A_1=0$ leads the doublets in $H_1$ to be massless,
However, the coupling $D_1 A H_2$ is also allowed.
Consequence of this coupling $D_1 A H_2$ is that the doublets in $H_2$ do not decouple from the other doublets
due to the mixing terms
\begin{eqnarray}
-\frac{i\sqrt{2}A_2}{3} (H_2^d D^{\prime u}-D^{\prime d} H_2^u)\nonumber
\end{eqnarray}
in the mass matrix,  which leaves no massless states in the weak doublets.
Thus direct application of the DW mechanism in the presence of  $\mathbf{120}$ does not
realize natural DTS.

\begin{figure}\begin{center}
  \vspace{-12cm}
  \includegraphics[width=20cm]{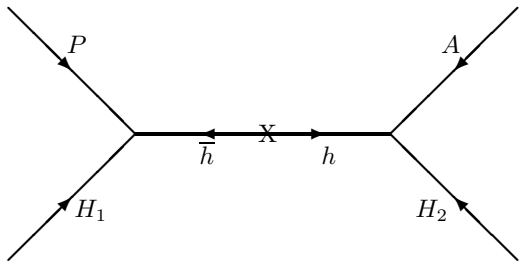}\\
  \vspace{-12cm}
  \caption{The filter sector which
  generates the $H_1 A (P) H_2$ coupling but forbids the $D_1 A (P) H_2$ coupling.}\label{fig}\end{center}
\end{figure}

To suppress the coupling $D_1 A H_2$ when $H_1$ and $D_1$ have the same symmetry behaviors under
the extra symmetry, we introduce a pair of $\mathbf{10}$s ($h+\overline{h}$) and a singlet ($P$) in SO(10)
and  replace $\frac{1}{2} m_H H_2^2 (+ S^{\prime} H_2^2) + H_1 A H_2$ in (\ref{WDT}) by
the following superpotential
\begin{eqnarray}\label{filt1}
W_{filter}= H_1 P \overline{h} + m_h \overline{h} h + h A H_2 + \frac{1}{2}m_2 H_2^2.
\end{eqnarray}
The mass matrix for the doublets and triplets are
\begin{eqnarray}\label{filter1}
M_D=\left(
\begin{array}{cccc}
0 & P & 0 & 0 \\
P & 0 &m_h& 0 \\
0 &m_h& 0 &A_1\\
0 & 0 &A_1&m_2
\end{array}
\right)
\end{eqnarray}
and
\begin{eqnarray}\label{filter11}
M_T=\left(
\begin{array}{cccc}
0 & P & 0 & 0 \\
P & 0 &m_h& 0 \\
0 &m_h& 0 &A_2\\
0 & 0 &A_2&m_2
\end{array}
\right),
\end{eqnarray}
respectively.
The orderings in both the columns and the rows are $(H_1, \overline{h}, h, H_2$) in (\ref{filter1},\ref{filter11}).
It is now clear if the DW solutions $A_1=0$ and $A_2\neq 0$ are taken,
there is a massless eigenvalue in (\ref{filter1}) with the eigenstates
\begin{eqnarray}\label{doub1}
H^u&=&\frac{m_h}{\sqrt{P^2+m_h^2}} H_1^u- \frac{P}{\sqrt{P^2+m_h^2}} h^u,\nonumber\\
H^d&=&\frac{m_h}{\sqrt{P^2+m_h^2}} H_1^d- \frac{P}{\sqrt{P^2+m_h^2}} h^d,
\end{eqnarray}
while no massless eigenvalue exists in the matrix $M_T$.
This realization of the DW mechanism of giving masses to the triplets can be depicted in Fig. \ref{fig}.
The coupling $D_1 A H_2$ is forbidden in the same time when the direct coupling $H_1 A H_2$ is absent.
Here the singlet $P$ plays the role as a ``filter'' which forbids
the unwanted coupling.

The situation does not change much when the massless eigenstates are got from the last section and are
not purely from $H_1$.
Supposing that these states are
\begin{eqnarray}
H_u^0&=&C_{H_1}^u H_1^u+\cdots, ~~~~H_d^0=C_{H_1}^d H_1^d+\cdots,
\nonumber\\
T^0&=&D_{H_1}^T H_1^T+\cdots, ~~~~\overline{T}^0=D_{H_1}^{\overline{T}} H_1^{\overline{T}}+\cdots,
\end{eqnarray}
(\ref{filter1}) and (\ref{filter11}) are modified into
\begin{eqnarray}\label{filter2}
M_D=\left(
\begin{array}{cccc}
0 & C_{H_1}^d P & 0 & 0 \\
C_{H_1}^u P & 0 &m_h& 0 \\
0 &m_h& 0 &A_1\\
0 & 0 &A_1&m_2
\end{array}
\right)
\end{eqnarray}
and
\begin{eqnarray}\label{filter21}
M_T=\left(
\begin{array}{cccc}
0 & D_{H_1}^{\overline{T}} P & 0 & 0 \\
D_{H_1}^T P & 0 &m_h& 0 \\
0 &m_h& 0 &A_2\\
0 & 0 &A_2&m_2
\end{array}
\right),
\end{eqnarray}
respectively.
Consequently, the conclusion which follows (\ref{filter1}) still holds,
provided that the massless doublets of the MSSM are now
\begin{eqnarray}\label{doub2}
H^u&=&\frac{m_h}{\sqrt{|C_{H_1}^u P|^2+m_h^2}}  H_u^0- \frac{C_{H_1}^u P }{\sqrt{|C_{H_1}^u P|^2+m_h^2}} h^u,\nonumber\\
H^d&=&\frac{m_h}{\sqrt{|C_{H_1}^d P|^2+m_h^2}}  H_d^0- \frac{C_{H_1}^d P}{\sqrt{|C_{H_1}^d P|^2+m_h^2}} h^d.
\end{eqnarray}
These contents of MSSM doublets can be also seen explicitly
in the full mass matrix of the doublets
\begin{eqnarray}\nonumber
\left(
\begin{array}{c|c|c}
\begin{array}{ccc} M_H & i\frac{A_1}{2}&\\-i\frac{A_1}{2}& &m_h\\&m_h&\end{array}
&\begin{array}{lccc}0&&&\\0&&&\\P~~~&&&\end{array}
&\\
\hline
\begin{array}{ccc}0~~~&0&~~~~P\\&&\\&&\\&&\end{array}
&0_{4\times 4}&\begin{array}{cc}A_{4\times3} &0_{4\times 1}\end{array}\\
\hline
\begin{array}{cccc}&&&\\&&&\\&&&\\&&&\end{array}&
B_{4\times 4}&C_{4\times 4}
\end{array}
\right),
\end{eqnarray}
where both the rows and the columns are ordered as\\
\centerline{
($H_2,h,\overline{h}; H_1,D_1,D_1^\prime,\overline{\Delta}; \Delta,D_2,D_2^\prime,\Phi$).}\\
It is clear that taking the DW solution $A_1=0$, a combination of $h^d,H_1^d,D_1^d,D_1^{\prime d},\overline{\Delta}^d$
gives the massless eigenstate $H_d$ of the MSSM,
and  a combination of  $h^u,H_1^u,D_1^u,D_1^{\prime u},\overline{\Delta}^u$ and ${\Phi}^u$ gives $H_u$.

The triplets  are relevant for proton decay amplitudes and their
treatments are different.
The triplets which do not couple with the fermions can be integrated out to get effective masses
for those triplets which couple with the fermions.
We can integrate out the triplets in $h,\overline{h},H_2$ in (\ref{filter11}) (not in (\ref{filter21})) firstly.
The result is generating an effective mass
\begin{eqnarray}\label{H1eff}
m_{H_1}^{eff}=\left(\frac{A_2}{m_h}\right)^2 \frac{P^2}{m_2}
\end{eqnarray}
for the triplets in $H_1$ responsible for proton decay.
To suppress proton decay mediated by the triplets in $H_1$, $m_{H_1}^{eff}$ needs to be enhanced.
Taking the mass parameters $m_h$ and $m_2$ the same order as the GUT scale determined by the GUT breaking VEVs
$\Phi_{1,2,3}\sim A_2\sim M_G$,
a large VEV $P\sim 10M_G$ is needed which may be realized through the Green-Shwarz mechanism \cite{green,green2,green3,green4}.
Note that in the original DW mechanism, the effective triplet mass is $\frac{A_2^2}{m_2}$
whose enhancement needs a small $m_2$ which, being the mass of the doublets of $H_2$, leads to
large splitting between the the doublets in $H_2$ and triplets in $H_{1,2}$, and the resultant threshold effect\cite{dwAAA,dwAAp}
spoils gauge coupling unification.
This problem is absent in the present model
since the threshold effect is proportional to \cite{zz}
\begin{eqnarray}\label{thres}
\ln \frac{M_G {\rm Det}^\prime(M_D)}{{\rm Det}(M_T)},
\end{eqnarray}
where ${\rm Det}^\prime(M_D)=
\lim_{\epsilon\rightarrow 0} \frac{1}{\epsilon}{\rm Det}(M_D+\epsilon {\rm\hat I}_{4\times 4})$\cite{msso10,msso102}.
For a large $P$, (\ref{thres}) is small so that gauge coupling unification is maintained.

The filter sector can be generalized beyond the superpotential (\ref{filt1}) if the
following terms
\begin{eqnarray}
m^\prime {\overline h}^2+m^{\prime\prime}\overline h H_2\nonumber
\end{eqnarray}
are added. They affect neither the massless eigenstates nor the threshold effects,
provided a large $P$ is taken.

There are more comments to be added if a filter sector is useful in building GUT models.
First, the singlet $P$ can be replaced by a $\mathbf{54}$ when it is needed.
Second, 
the easiest way to construct the filter sector superpotential (\ref{filter1}) is using
a new $\mathbb{Z}_2$ symmetry with the only fields odd under this $\mathbb{Z}_2$ symmetry are $P, h, \overline{h}$
and $H_2$.

\section{The full model}

Having all the requirements given,
we give a full model with DW mechanism to realize the DTS.
We impose a $\mathbb{Z}_{12}\times\mathbb{Z}_{2}$ symmetry.
The superfields and their transformation properties are summarized in Table II.

The most general superpotential for the Higgs sector is
\begin{eqnarray}\label{Wfull}
W&=&W_{DW}+W_{SB}+W_{filter}^\prime+W_{DT}^\prime 
\end{eqnarray}
where
$W_{DW}$ and $W_{SB}$ are given in (\ref{WDW}) and (\ref{WSB}), respectively,
\begin{eqnarray}
W_{filter}^\prime=W_{filter}+\overline{h} S^\prime h 
\end{eqnarray}
is the complete filter sector, and
\begin{eqnarray}
W_{DT}^\prime &=&(H_1 + D_1 +\overline{\Delta}) \Phi (\Delta  +  D_2)\nonumber\\
&+& m_D D_1 D_2 + m_\Delta \Delta\overline{\Delta} +D_1 S^\prime D_2\label{WDTp}
\end{eqnarray}
is the main doublet-triplet sector.

\begin{table}
\begin{tabular}{|c|c|c|c|c|c|c|c|c|c|c|c|c|c|c|c|c|c|}
  \hline
    & $H_1$ & $\overline{h}$ & $h$ & $H_2$ & $D_1$ & $D_2$ & $\overline{\Delta}$ & $\Delta$
    & $\Phi$ & $A$ & $A^\prime$ & $A^{\prime\prime}$& $S$ & $S^\prime$ & $P$ & $Y$\\
  \hline
  $\mathbb{Z}_{12}$ & 2 & 5  & -5 & 6 & 2 & -2 & 2 & -2 & 0 & -1 & -5 & 6 & 6 & 0 &5 &6
\\
  \hline
  $\mathbb{Z}_{2}$ & 0 & 1  & 1 & 1 & 0 & 0 & 0 & 0 & 0 & 0 & 0 & 0 & 0 & 0 &1 &0
\\\hline
\end{tabular}\label{tab2}
\caption{Fields and symmetric properties under $\mathbb{Z}_{12}\times\mathbb{Z}_{2}$
in the full model with $\mathbf{120}$ included. The charges for all the matter fields
are $(-1,0)$ under this symmetry.}
\end{table}

SUSY at high energy requires the F-flatness conditions.
Among them only $F_P=0$ is automatic so that the VEV $P$ can take a value larger
than the GUT scale through the Green-Schwarz mechanism\cite{green,green2,green3,green4},
as the $\mathbb{Z}_{12}\times\mathbb{Z}_{2}$ are embed into $U(1)$'s.
Since these U(1)'s are anomalous, the D-terms have the form
\be
D_A=-\xi+\sum Q_i|\phi_i|^2, ~~\xi=\frac{Tr ~Q}{192\pi^2} M_{Planck}^2,\nonumber
\ee
where $Q_i$ is the U(1) charge of the scalar field $\phi_i$, thus the VEV of $P$
is related the Planck scale $M_{Planck}$.
All the other VEVs are constrained by the F-flatness conditions
and have solutions of the GUT scale values.

The present model (\ref{Wfull}) has the following consequences.
First, it contains Higgs in $\mathbf{10}$, $\mathbf{\overline{126}}$ and $\mathbf{120}$ which
transform the same way under the extra symmetry so that they can give the masses and mixing
of the three generation matter particles through renormalizble Yukawa couplings.
Second, it realizes naturally the DTS through the DW mechanism supplied by a filter sector.
Third, proton decay amplitudes through dimension-five operators mediated by the Higgs triplets in $\mathbf{10}$
is suppressed without large threshold effect,
although general suppressions of all proton decay amplitudes 
are not discussed which need structural constructions of models like is done in \cite{dlz2}
and are beyond present study.
Finally, we need to mention that from (\ref{doub2}) that the up-type Higgs doublet of the MSSM
has a large component from $h$ which does not couple with the matter fields,
which might rise a difficulty in generating the large top quark mass whose Yukawa coupling is large.
In a more realistic model, we need to modify the VEV of $P$ by a GUT-valued VEV of
a SO(10) singlet or $\mathbf{54}$ to fix this problem. To suppress  proton decay mediated by
the color triplets in $H_1$, a small $m_h$ needs to be taken whose origin might be related
to the seesaw scale $M_{seesaw}\sim \frac{M_{GUT}^2}{M_{Planck}}$
generated through the Green-Scwarz mechanism,
meanwhile all the other proton decay amplitudes are also suppressed by  $\frac{M_{seesaw}}{M_{GUT}}$ \cite{LiZhang}.

\section{Summary}

In the present work we have proposed a renormalizable SUSY SO(10)
model of naturally realized DTS through the DW mechanism.
A filter sector is introduced to forbid  an unwanted coupling which spoils the DW mechanism.
Proton decay mediated by the color-triplets in $H_1$
can be suppressed without spoiling gauge unification. However, a problem on the contents of
the MSSM Higgs doublets which are insufficient to give top quark correct mass.
This problem requires the present model to be further improved.

As in all renormalizable SUSY GUT models,
the large representations used in the present model also bring in
big contributions to the $\beta$-function of the GUT gauge coupling.
Consequently,
the GUT gauge coupling blows up quickly above the GUT scale.
This non-perturbative problem may not be a serious one if we take the
following picture. The universe was in the GUT symmetric phase at very high temperature in its early stage.
As the universe was cooling down, phase transition happened and GUT symmetry was broken.
Without knowing more details about what was happening during this phase transition,
the non-perturbative behavior of the GUT gauge coupling above the GUT scale  may not be
a problem.

\section*{Appendix}
Denoting
\bea
m_{2,3}=m_{3,2}&=&m_h+\frac{1}{2} \sqrt{\frac{3}{5}} S', \nonumber\\
m_{4,7}&=&\frac{\Phi _2}{\sqrt{10}}-\frac{\Phi _3}{2 \sqrt{5}}, \nonumber\\
m_{5,9}=m_{9,5}&=&m_d+\frac{1}{2} \sqrt{\frac{3}{5}} S', \nonumber\\
m_{6,7}&=&\frac{\Phi _1}{4 \sqrt{15}}-\frac{\Phi _3}{6 \sqrt{10}}, \nonumber\\
m_{6,10}=m_{10,6}&=&m_d-\frac{S'}{6 \sqrt{15}}+\frac{\sqrt{2} \Phi _2}{9}, \nonumber\\
m_{7,4}&=&-\frac{\Phi _2}{\sqrt{10}}-\frac{\Phi _3}{2 \sqrt{5}}, \nonumber\\
m_{7,6}&=&\frac{\Phi _1}{4 \sqrt{15}}+\frac{\Phi _3}{6 \sqrt{10}}, \nonumber\\
m_{7,8}&=&m_{\Delta }+\frac{\Phi _2}{15 \sqrt{2}}+\frac{\Phi _3}{30}, \nonumber\\
m_{8,7}&=&m_{\Delta }+\frac{\Phi _2}{15 \sqrt{2}}-\frac{\Phi _3}{30}, \nonumber\\
m_{8,10}&=&\frac{\Phi _1}{4 \sqrt{15}}-\frac{\Phi _3}{6 \sqrt{10}}, \nonumber\\
m_{10,8}&=&\frac{\Phi _1}{4 \sqrt{15}}+\frac{\Phi _3}{6 \sqrt{10}}, \nonumber\\
m_{11}&=&m_{\Phi }-\frac{1}{4} \sqrt{\frac{3}{5}} S'+\frac{\Phi _2}{\sqrt{2}}+\frac{\Phi _3}{2}, \nonumber
\eea
and ordering both the rows and the columns as\\
\centerline{($H_2,h,\overline{h}; H_1,D_1,D_1^\prime,\overline{\Delta}; \Delta,D_2,D_2^\prime,\Phi$),}\\
the mass matrix for all the weak doublets in the
full model can be given after the most general superpotential (5.1),

\newpage
\vspace*{-1cm}
$\left(\begin{array}{ccccccccccc}
 m_2 & \frac{i A_1}{2} & 0 & 0 & 0 & 0 & 0 & 0 & 0 & 0 & 0 \\
 -\frac{i A_1}{2} & 0 & m_{2,3} & 0 & 0 & 0 & 0 & 0 & 0 & 0 & 0 \\
 0 & m_{3,2} & 0 & P & 0 & 0 & 0 & 0 & 0 & 0 & 0 \\
 0 & 0 & P & 0 & 0 & 0 & 0 & m_{4,8} & -\frac{\Phi _1}{2} & -\frac{\Phi _3}{2 \sqrt{2}} & 0 \\
 0 & 0 & 0 & 0 & 0 & 0 & 0 & \frac{\Phi _3}{4 \sqrt{30}} & m_{5,9} & \frac{\Phi _3}{6 \sqrt{3}} & 0 \\
 0 & 0 & 0 & 0 & 0 & 0 & 0 & m_{6,8} & \frac{\Phi _3}{6 \sqrt{3}} & m_{6,10} & 0 \\
 0 & 0 & 0 & 0 & 0 & 0 & 0 & m_{7,8} & \frac{\Phi _3}{4 \sqrt{30}} & m_{7,10} & 0 \\
 0 & 0 & 0 & m_{8,4} & \frac{\Phi _3}{4 \sqrt{30}} & m_{8,6} & m_{8,7} & 0 & 0 & 0 & \frac{\overline{v_R}}{10} \\
 0 & 0 & 0 & -\frac{\Phi _1}{2} & m_{9,5} & \frac{\Phi _3}{6 \sqrt{3}} & \frac{\Phi _3}{4 \sqrt{30}} & 0 & 0 & 0 & -\frac{\overline{v_R}}{2 \sqrt{30}} \\
 0 & 0 & 0 & -\frac{\Phi _3}{2 \sqrt{2}} & \frac{\Phi _3}{6 \sqrt{3}} & m_{10,6} & m_{10,7} & 0 & 0 & 0 & -\frac{\overline{v_R}}{2 \sqrt{10}} \\
 0 & 0 & 0 & -\frac{v_R}{\sqrt{5}} & -\frac{v_R}{2 \sqrt{30}} & -\frac{v_R}{2 \sqrt{10}} & \frac{v_R}{10} & 0 & 0 & 0 & m_{11}
\end{array}\right).$


\end{document}